\documentclass[prb,twocolumn,showpacs,preprintnumbers,amsmath,amssymb]{revtex4}
\usepackage{amsmath,amssymb,amsfonts} % Typical maths resource packages
\usepackage{graphics}                 % Packages to allow inclusion of graphics
\usepackage[dvips]{graphicx}
\usepackage[latin1]{inputenc}

\begin{document}

\title{Hot electron mediated desorption rates calculated from excited state potential energy surfaces}

\author{Thomas Olsen}
\author{Jeppe Gavnholt}
\author{Jakob Schi\o tz}
\email{schiotz@fysik.dtu.dk}

\affiliation{Danish National Research Foundation's Center of Individual Nanoparticle Functionality (CINF),
	Department of Physics,
	Technical University of Denmark,
	DK--2800 Kongens Lyngby,
	Denmark}%

\date{\today}% It is always \today, today,
             %  but any date may be explicitly specified

\begin{abstract}
We present a model for Desorption Induced by (Multiple) Electronic Transitions (DIET/DIMET) based on potential energy surfaces calculated with the Delta Self-Consistent Field extension of Density Functional Theory. We calculate potential energy surfaces of CO and NO molecules adsorbed on various transition metal surfaces, and show that classical nuclear dynamics does not suffice for propagation in the excited state. We present a simple Hamiltonian describing the system, with parameters obtained from the excited state potential energy surface, and show that this model can describe desorption dynamics in both the DIET and DIMET regime, and reproduce the power law behavior observed experimentally. We observe that the internal stretch degree of freedom in the molecules is crucial for the energy transfer between the hot electrons and the molecule when the coupling to the surface is strong.

\end{abstract}
\pacs{31.15.xr, 71.15.Qe, 71.38.-k, 82.20.Gk, 82.20.Kh}
\maketitle

\section{Introduction}
The advent of femtosecond lasers has initiated major progress in the study of non-adiabatic surface dynamics on a wide range of systems. Photo-induced desorption had already been observed for a few adsorbate systems\cite{buntin1,buntin2} using low-intensity nanosecond laser pulses, but high-intensity femtosecond laser pulses have been shown to induce desorption in a large class of adsorbate systems\cite{prybyla1,prybyla2,budde,misewich2,struck,howe,cai,fournier} and induce chemical reactions which cannot proceed by thermal heating.\cite{bonn2}

The mechanism attributed to these reactions is excitation of substrate electrons by the laser pulse. A single hot electrons can then interact with an initially unoccupied adsorbate resonance thus asserting a force on the adsorbate nuclei which may then lead to Desorption Induced by Electronic Transitions (DIET). Using femtosecond lasers it is possible to reach high densities of excited electrons resulting in a new dominating mechanism - Desorption Induced by Multiple Electronic Transitions (DIMET)\cite{misewich1} where several hot electrons interact with the adsorbate. 

A different method to produce hot electron based on a Metal-Insulator-Metal (MIM) heterostructure was suggested by Gadzuk\cite{gadzuk96}. With an ideal MIM device it is possible to tune hot electrons to any desired resonance of an adsorbate system and the approach thereby suggests the highly attractive possibility of performing selective chemistry at surfaces. Such devices have been constructed and characterized\cite{thomsen} and comprise a promising candidate for future hot electron femtochemistry experiments. 
 
The theoretical framework to describe the non-adiabatic dynamics resulting from a hot electron interacting with an adsorbate is usually based on the concept of Potential Energy Surfaces (PES). In the Born-Oppenheimer approximation the electrons are assumed to remain in their ground state and are thus decoupled from the nuclei. This allow one to map out a ground state PES for the nuclei by calculating the electronic energy for each position of the nuclei. Similarly, when an initially unoccupied resonance becomes occupied a new excited state PES arises which has its minimum at a different location than the ground state PES and a force is exerted on the adsorbate. Several models have emerged to deal with non-adiabatic dynamics at surfaces but they are usually limited by the difficulty to obtain reliable excited state PESs and most theoretical results are based on model potentials.\cite{gadzuk90,gadzuk91,gadzuk95,harris,gadzuk97,saalfrank}

An often used method to treat the extreme DIMET regime with many contributing electrons is using an electronic friction model.\cite{nourtier,newns,brandbyge} The hot electrons are then assumed to thermalize rapidly and the influence of the electrons on the adsorbate are treated statistically using an electronic temperature which can be several thousand Kelvin. The conceptual picture is that of a hot Fermi distribution with a tail partially overlapping an adsorbate resonance and thereby exerting a force on the adsorbate. However, correct calculation of the temperature dependent friction still requires knowledge of the excited state PES.

The subject of this paper will be the application of two-dimensional excited state PESs to calculate desorption probabilities. We will be particularly interested in the DIET regime where the hot electron has a known energy as relevant for the MIM device and the few-electron DIMET regime. Although the friction models have enjoyed some success\cite{luntz1,luntz2}, there is still a need of a microscopic non-statistical model of DIMET to test the assumption of thermally equilibrated electrons and to bridge the gap to the DIET regime. Furthermore, the hot electron femtochemistry relevant to the MIM device can certainly not be described using an electronic temperature since all electrons are tuned to a specific energy.

We start by summarizing the method of linear expansion $\Delta$SCF-DFT\cite{gavnholt} used to calculate the excited state PESs and note some qualitative features using CO on Pt(111) as an example. We then discuss the models used to obtain desorption probabilities based on the calculated potential energy surfaces. First an adiabatic model in which the adsorbate jumps between the ground and excited state potentials is presented. A general non-adiabatic Newns-Anderson like model\cite{newnsanderson1,newnsanderson2} is then introduced and the connection to potential energy surfaces is explained. This model with linear coupling has previously been solved\cite{wingreen} and applied to the one-dimensional desorption problem with model parameters\cite{gadzuk91}. We extend these results to a two-dimensional adsorbate and obtain the non-adiabatic coupling parameters from calculated excited state potential energy surfaces. In the DIET regime the model will be used to show that for small excited state lifetimes the main channel of energy transfer is the internal degree of freedom, and we emphasize its importance in desorption dynamics. We compare the calculated desorption probabilities for CO and NO on four transition metal surfaces and note some general features of the desorption dynamics. The scattering probabilities obtained in the model are then generalized to include adsorbates in any vibrationally excited state which allow us to extend the calculations to include a substrate temperature and to treat the DIMET regime within the model. In the appendix it is shown how to expand excited states within the projector augmented wave formalism and the results and generalizations of scattering amplitude calculations are summarized.

\section{Potential energy surfaces}
The potential energy surfaces were obtained using the code
\texttt{gpaw}\cite{gpaw, mortensen} which is a real-space Density
Functional Theory (DFT) code that uses the projector augmented wave
method.\cite{blochl1,blochl2} In all our calculations we used the
Revised Perdew-Burke-Ernzerhof (RPBE) exchange-correlation functional
\cite{hammer} since this has been designed to perform well for
molecules adsorbed on surfaces, and has been shown to perform better
than the original PBE functional\cite{PBE} both for isolated
molecules \cite{kurth} and for adsorbed molecules{hammer}. For each
metal we set up a closed-packed surface consisting of three atomic
layers with the top layer being relaxed. 10 \AA\ of vacuum was then
introduced above the slab and 0.25 monolayer of adsorbate molecules
relaxed at either top or hcp hollow site. We then mapped out
two-dimensional ground state potential energy surfaces in terms of the
internal stretch and the center of mass (COM) to surface distance
coordinate using 12 irreducible k-points and a grid spacing of 0.2
\AA.

To find the excited state potential energy surfaces we use the method
of linear expansion $\Delta$SCF which we have published in a previous
work\cite{gavnholt} and implemented in \texttt{gpaw}.  In the previous
publication we have tested the method against inverse
photo-emission spectroscopy, and found that it performed well for
molecules chemisorbed on surfaces.\cite{gavnholt} In each step of the
self consistency cycle an electron is removed from the Fermi level,
the density of an excited state is added to the total density, and the
band energy of this state is added to the total energy. To get the
band energy right we need to expand the excited state on the Kohn-Sham
orbitals found in each iteration. The method is thus a generalization
of the usual $\Delta$SCF where occupations numbers are
changed. Instead of changing occupation numbers we occupy an orbital
which is not an eigenstate of the KS Hamiltonian but a superposition
of eigenstates in such a way that the state is as close as possible to
the original molecular state.  We refer to appendix \ref{paw} for
details on how to do this within the projector augmented wave
formalism. The excited
states used in this paper are the anti-bonding $2\pi$ orbitals of NO
and CO.

In the previous publication,\cite{gavnholt} we investigated the
influence of the interactions between neighboring super
cells for different super-cell sizes, and found that the size-dependency
of the excitation energy is consistent with an electrostatic
dipole-dipole interaction.  Already for a $(2\times2)$ surface cell,
the interaction energy has become small, and furthermore this
interaction energy will have little influence on the slope of the
excited-state PES, and thus little influence on the calculated
desorption rates.  For this reason, and to keep the calculations
manageable, we use a $(2\times2)$ surface cell. 

\begin{figure}[t]
	\includegraphics[scale=0.40]{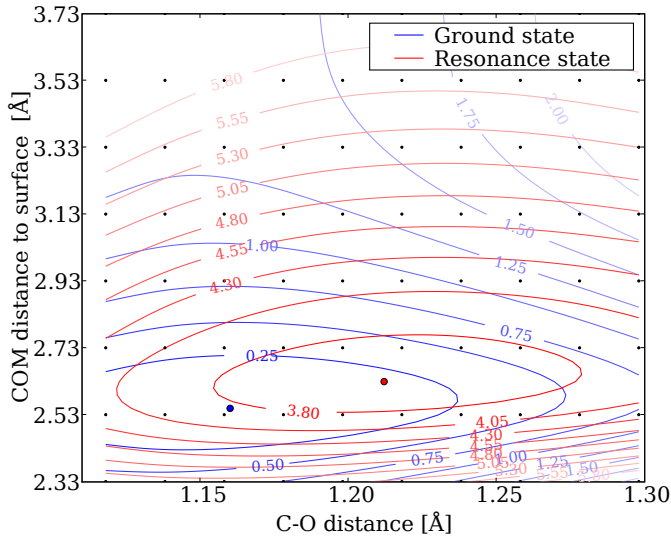} 
\caption{Ground and excited state ($2\pi$) potential energy surfaces for CO adsorbed on Pt(111) - top site. The coverage is 0.25 monolayer.}
\label{fig:pes}
 \end{figure}

As an example we show the two-dimensional excited state PES superimposed on a ground state PES in the case of CO on Pt(111) top site in Fig. \ref{fig:pes}. The molecules adsorb with the molecular axis perpendicular to the surface with carbon closest to the top site. Due to the symmetry of the $2\pi$ orbital and the geometry at the ground state minimum we cannot induce forces parallel to the surface if the molecule is at the ground state minimum when excited. The excited state could have unstable extremal points with respect to the degrees of freedom parallel to the surface, but the model we apply in this work only depend on the degrees of freedom with non-vanishing derivatives on the excited state PES and we thus assume that the center of mass (COM) and internal stretch degrees of freedom should capture the essential desorption dynamics of the considered systems.

Since the excited molecule has an extra electron in an anti-bonding orbital the excited molecule is expected to have a larger equilibrium bond length and this is also what we observe. A popular and conceptually simple way of explaining desorption in one-dimensional models of DIET is the Antoniewicz mechanism,\cite{antoniewicz} where the excited molecule induces an image charge on the surface which results in an attractive force on the surface. The excited molecule is then accelerated towards the surface and eventually decays to the steep wall of the ground state Morse potential. From Fig. \ref{fig:pes} we observe a qualitatively different behavior: the COM of the excited molecules experience a repulsive force accelerating the COM of the molecule away from the surface. This is due to the effect of the bond length expansion and the fact that the $2\pi$ orbital have a large density in the vicinity of the carbon atom which gives a repulsion that dominates the image charge attraction. It will be shown below that for the considered systems it is primarily excitation of the internal degree of freedom which is responsible for the large energy transfers leading to desorption.

The potential energy surfaces for CO adsorbed on Pd, Rh and Ru show very similar qualitative features. 

\section{Models}
The timescale at which adsorbates dissipate energy to the substrate is typically on the order of picoseconds\cite{beckerle} and since the oscillation times for the two modes is $\sim 15-50\;fs$ (see section \ref{sec:par}) we will assume that the molecule has plenty of time to desorb if it has absorbed the required energy from a hot electron. This is the major assumption we will impose and thus when we refer to desorption rates in the following it is the rate of transferring at least of the energy needed for a molecule to desorb.

Assuming a Lorentzian resonance with Full Width at Half Maximum (FWHM) $\Gamma=\hbar/\tau$ centered at $\epsilon_a$, the probability that a hot electron of energy $\varepsilon$ desorbs the molecule becomes
\begin{equation}\label{clas}
P_{Des}^{Ad}(\varepsilon)=\frac{(\Gamma/2)^2}{(\varepsilon-\epsilon_a)^2+(\Gamma/2)^2}\frac{1}{\tau}\int_{0}
^\infty P(t_d)e^{-t_d/\tau}dt_d,
\end{equation}
where $P(t_d)$ is the probability of a desorption event when the molecule is excited at $t=0$ and decays at time $t=t_d$.
Using classical dynamics the probability $P(t_d)$ can be obtained by propagating the molecule on the excited state PES according to the forces, evaluate the energy gain $\Delta E$ after time $t_d$ and put $P(t_d)=1$ if $\Delta E>E_{Des}$ and $P(t_d)=0$ if $\Delta E<E_{Des}$.
However, the short lifetime of the excited electron implies that classical molecular propagation on the excited state PES may not 
be a good approximation. 

In fact, the classical limit is obtained when the action $S=\int dtL(\dot{x}(t),x(t))$ on a representative path satisfies
\begin{equation}
|S|\gg\hbar.
\end{equation}
Assuming a quadratic excited state potential of frequency $\omega$ and initial potential energy $E_0$ we can evaluate 
the action on a classical path between initial time $t_i$ and final time $t_f$. For generic timescales one just 
obtains the usual condition of high excitation numbers $E_0\gg\hbar\omega$, whereas 
for $\omega\Delta t\ll 1$, the additional condition of $E_0\Delta t\gg\hbar$ needs to be satisfied in order to apply classical dynamics. In the case of CO on Pt(111) we have 
$E_0\sim0.3\;eV$ (Fig. \ref{fig:pes}) and $\tau\sim1\;fs$ (Fig. \ref{fig:pdos}) which gives $E_0\Delta t\sim\hbar$. Thus molecular propagation on 
the excited state PES is not expected to follow the classical equations of motion. Below we will show an example where a classical analysis underestimates desorption probabilities by several orders of magnitude.

This scheme could be extended to a quantum dynamical treatment of the molecule by
propagating the molecular wavefunction using a two-PES Hamiltonian.
However, the method still rests on the Born-Oppenheimer approximation and the adiabatic concept of potential energy surfaces and thus cannot be expected to fully capture the non-adiabatic entangled dynamics of the resonant electron and adsorbate coordinates. 

Instead we consider a Newns-Anderson \cite{newnsanderson1,newnsanderson2} type Hamiltonian with substrate states $|k\rangle$, a resonant state $|a\rangle$, adsorbate coordinates $x_i$, an adiabatic adsorbate ground state potential $V_0(x_i)$, and a non-adiabatic coupling of the resonant electron to adsorbate coordinates $\varepsilon_a(x_i)$
\begin{align}\label{h_non-ad}
H&=T(\dot{x}_i)+V_0(x_i)+\varepsilon_a(x_i)c_a^{\dag}c_a\notag\\
&+\sum_k\epsilon_kc_k^{\dag}c_k+\sum_k\Big(V_{ak}c_a^{\dag}c_k+V_{ak}^*c_k^{\dag}c_a\Big).
\end{align}
The strength of the electronic coupling is expressed through the function:
\begin{equation}\label{gamma}
 \Gamma(\varepsilon)=2\pi\sum_k|V_{ak}|^2\delta(\varepsilon-\epsilon_k).
\end{equation}
The model as such neglects the electron-electron interaction, but we assume that the important part of the electron-electron interactions is the restructuring of the metallic electrons when the resonance is occupied and that we can capture this effect in an effective non-adiabatic coupling. To do this we note that we can obtain $\varepsilon_a(x_i)$ as the expectation value differences of \eqref{h_non-ad} with the adsorbate at $x_i$ with and without an electron in the state $|a\rangle$. Applying this to an interacting problem lead us to identify $\varepsilon_a(x_i)=V_1(x_i)-V_0(x_i)$ where $V_1(x_i)$ and $V_0(x_i)$ are the potential energy surfaces of excited and ground states which we have obtained with linear expansion $\Delta$SCF-DFT.

In the following we will apply the wide band limit which means that the individual coupling coefficients $V_{ak}$ are assumed to vary slowly in energy and the density of states $\rho(\varepsilon)$ is taken as constant in the vicinity of the resonance. This gives an energy independent coupling $\Gamma=2\pi\rho(\epsilon_a)\sum_k|V_{ak}|^2$ and the resonance spectral function corresponding to the electronic part of \eqref{h_non-ad} becomes a Lorentzian with FWHM $\Gamma$.
 
Even in the wide band limit it is quite difficult to handle the model \eqref{h_non-ad} analytically with arbitrary coupling function $\varepsilon_a(x_i)$. In particular, we would like to calculate the probability that an incoming substrate electron of energy $\varepsilon_i$ scatters inelastically on the resonance and is reflected back into the substrate with final energy $\varepsilon_f$.
Fortunately, the potential energy surfaces we are considering are close to being quadratic in the region of interest (see Fig. \ref{fig:pes}) and the ground and excited state potentials have approximately the same curvature.
Taylor expanding $V_0(x_i)$ to second order and $\varepsilon_a(x_i)$ to first order in the vicinity of the ground state equilibrium positions $x_i^0$ then gives
\begin{align}\label{h_quadratic}
H&=\epsilon_ac_a^{\dag}c_a + \sum_k\epsilon_kc_k^{\dag}c_k+\sum_k\Big(V_{ak}c_a^{\dag}c_k+V_{ak}^*c_k^{\dag}c_a\Big)\notag\\
&+\sum_i\hbar\omega_i(a_i^{\dag}a_i+\frac{1}{2})+\sum_i\lambda_ic_a^{\dag}c_a(a_i^{\dag}+a_i),
\end{align}
with $\epsilon_a=V_1(x_i^0)-V_0(x_i^0)$ and
\begin{align}
\lambda_i=\frac{l_i}{\sqrt{2}}\frac{\partial}{\partial x_i}V_1\Big|_{x_i=x_i^0}, \qquad l_i=\sqrt{\frac{\hbar}{m_i\omega_i}},
\end{align}
where we have assumed that an appropriate transformation to normal coordinates has been performed. Note that if the ground and excited state potentials are exactly quadratic with equal second derivatives, we can relate the coupling constants to the positions $x_i^1$ of the excited state potential minimum as $\lambda_i^2=\hbar\omega_i\Delta V_i$ with $\Delta V_i=\frac{1}{2}m_i\omega_i^2|x_i^1-x_i^0|$. The quantity $g_i=(\lambda_i/\hbar\omega_i)^2$ then corresponds to an ''initial quantum number'' on the excited state surface and this becomes the effective dimensionless coupling constant in the model (see appendix \ref{prob}). The Hamiltonian \eqref{h_quadratic} has previously been subject to detailed analysis in the context of inelastic scattering\cite{wingreen} and applied to desorption dynamics \cite{gadzuk91} for the case of a one-dimensional adsorbate with model parameters. 

Below we extend the results of Refs. \onlinecite{gadzuk91} and \onlinecite{wingreen} to a two-dimensional adsorbate and calculated the coupling parameters $\lambda_i$ from excited state potential energy surfaces. We also calculate scattering amplitudes for an adsorbate initially in an vibrationally excited state which enable us to apply the model to the DIMET regime.

\subsubsection{DIET}
In Eq. \eqref{P_mn} we show how to calculate the scattering probability $P_{n_i,n_j}(\varepsilon_i)$ that an incoming electron of energy $\varepsilon_i$ excites the $(n_i,n_j)$ mode of a two-dimensional harmonic oscillator. The probability of transferring $E_R$ or more energy to the adsorbate can then be found by calculating
\begin{equation}\label{diet}
 P_R(\varepsilon_i) = \sum_{n_i,n_j} P_{n_in_j}(\varepsilon_i)\theta(\hbar\omega_in_i+\hbar\omega_jn_j-E_R),
\end{equation}
where $\theta(x)$ is the Heaviside step function. The desorption rate can then be calculated by integrating this expression with the current density of incoming hot electrons.
One should note that the probability $P_{n_in_j}$ of exciting the $(n_i,n_j)$ modes in a two-mode model is not just given by the product of single mode probabilities $P_n$ in a one-mode model. This is due to an indirect coupling of the two modes through the resonance.
The result can be generalized to include the substrate temperature and we will examine the consequences of this below. 

\subsubsection{DIMET}
If we assume that the time between individual inelastic scattering events is much longer than the scattering time itself, it is possible to regard multiple-electron desorption events as a sequence of single-electron scattering events. Since we have extended the inelastic scattering probabilities to include situations where the molecule is initially in a vibrationally excited state, it is also possible to treat DIMET events within the model \eqref{h_quadratic}. As an example, let us assume a single vibrational mode which is initially unoccupied ($n=0$). When a hot electron with energy $\varepsilon_1$ scatters inelastically on the resonance the result will be a probability distribution $P_{n_1}(\varepsilon_1)$ for all vibrationally excited states $n$ of the molecule. If a second electron with energy $\varepsilon_2$ now scatters on the resonance the probability distribution will change to $P_{n_2}(\varepsilon_1,\varepsilon_2)$ and so forth. The probability $P_{n_2-n_1}(\varepsilon_2, n_1)$ of exciting the state $n_2$ given that the initial state were $n_1$ is calculated in Eq. \eqref{P_n} and we can write
\begin{equation}
 P_{n_2}(\varepsilon_1,\varepsilon_2)=\sum_{n_1=0}^{\infty}P_{n_2-n_1}(\varepsilon_2, n_1)P_{n_1}(\varepsilon_1)
\end{equation}
for a two-electron event and similar expressions for multiple-electron events. Given an initial distribution of hot electrons we may then calculate the probability of a desorption event with any number of contributing electrons.

\section{Results}
\subsection{Parameters}\label{sec:par}
The parameters in the desorption model \eqref{h_quadratic} are the width of the resonance $\Gamma$, the frequencies of the normal modes $\omega_i$, the excitation energy $\epsilon_a$ and the non-adiabatic coupling coefficients $\lambda_i$. 
We cannot calculate $\Gamma$ from first principles but we estimate its value from the Kohn-Sham projected density of states. It is typically on the order of $1\;eV$, but it will be instructive to treat it as a free parameter and examine how it affects desorption probabilities. 

The frequencies are obtained from a standard normal mode analysis and $\epsilon_a$ is obtained as the excitation energy at the ground state potential minimum. The coupling coefficients are determined by mapping out a small area of the excited state potential energy surface in the immediate vicinity of the ground state potential. In each of the considered systems we optimize the area such that it is small enough to be linear but large enough to suppress numerical fluctuations in the excited state energies. We then fit a linear function to this area and transform the derivatives to the normal modes. 

In all the considered systems the calculated normal modes are similar but not identical to the standard COM and internal stretch modes.
For example with CO on Pt(111) the internal stretch and COM modes are respectively: $\mathbf{d}=(-1,0.75)$ and $\mathbf{z}=(1,1)$ whereas the calculated modes are in the directions $\mathbf{d}=(-1,0.68)$ and $\mathbf{z}=(1,1.11)$ with respect to the $(x_C,x_O)$ coordinates normal to the surface. 
Since the desorption probabilities are quite sensitive to the value of the non-adiabatic coupling constants it is important that we take the derivatives on the excited state PES with respect to the correct normal modes.

Tables \ref{tab:par1} and \ref{tab:par2} below display the calculated parameters. We have only examined CO at on-top sites and NO at hcp hollow sites. NO is seen to have much lower non-adiabatic coupling coefficients and excitation energies than CO. The low excitation energies is due to the fact that NO already has one electron in the anti-bonding orbital and the resonance thus has to lie close to the Fermi level of the metal. The small coupling coefficients can also be traced to the ground state occupation of the $2\pi$ orbital on NO. In the Kohn-Sham picture we can imagine the resonance corresponding to $2\pi$ lying right at the Fermi level being partially occupied. When an extra electron is put into the orbital the resonance energy is increased due to Hartree repulsion and the initial partial occupation is lost. In the true system things are more complicated, but the qualitative features are the same: exciting NO results in less charge being transferred to the molecule than exciting CO and thus a weaker non-adiabatic coupling. 
Thus it is much harder to transfer energy to adsorbed NO compared to CO in a one electron event, but since the resonance is located much closer to the Fermi level a thermal distribution of hot electrons is likely to result in more frequent scattering events than for CO.
\begin{table}[t]
\begin{center}
\begin{tabular}{c|c|cc|cc}
Metal &  $\epsilon_a$ & $\omega_z$ & $\omega_d$  & $\lambda_z$ & $\lambda_d$  \\
	\hline
Pt(111)  & 3.89 & 0.054 & 0.255  & -0.142 & -0.145 \\
Pd(111)  & 3.64 & 0.061 & 0.256  & -0.082 & -0.164 \\
Rh(111)  & 3.80 & 0.048 & 0.247  & -0.129 & -0.132 \\
Ru(0001) & 3.74 & 0.054 & 0.255  & -0.134 & -0.120 \\
\end{tabular}
\end{center}
\caption{Parameters for CO adsorbed at top site on four transition metals. All number are $eV$.}
\label{tab:par1}
\end{table}
\begin{table}[t]
\begin{center}
\begin{tabular}{c|c|cc|cc}
Metal &  $\epsilon_a$ & $\omega_z$ & $\omega_d$  & $\lambda_z$ & $\lambda_d$  \\
	\hline
Pt(111)  & 1.71 & 0.039 & 0.196  & -0.050 & -0.053 \\
Pd(111)  & 1.48 & 0.055 & 0.201  & -0.046 & -0.053 \\
Rh(111)  & 1.82 & 0.073 & 0.277  & -0.042 & -0.020 \\
Ru(0001) & 2.14 & 0.042 & 0.192  & -0.052 & -0.006 \\
\end{tabular}
\end{center}
\caption{Parameters for NO adsorbed at hcp hollow site on four transition metals. All number are $eV$.}
\label{tab:par2}
\end{table}

\subsection{DIET desorption rates}
The probability that a single electron with energy $\varepsilon_i$ scatters inelastically and transfers the energy $E_R$ to an adsorbate can be calculated in the model \eqref{h_quadratic} with Eq. \eqref{diet}. Our basic assumption is that rate of energy dissipation to the substrate is much longer than the time of a desorption event and when we refer to desorption rates in the following it will mean the rates of transferring the energy needed for a molecule to desorb in a truncated quadratic potential.

\begin{figure}[t]
  	\includegraphics[scale=0.4]{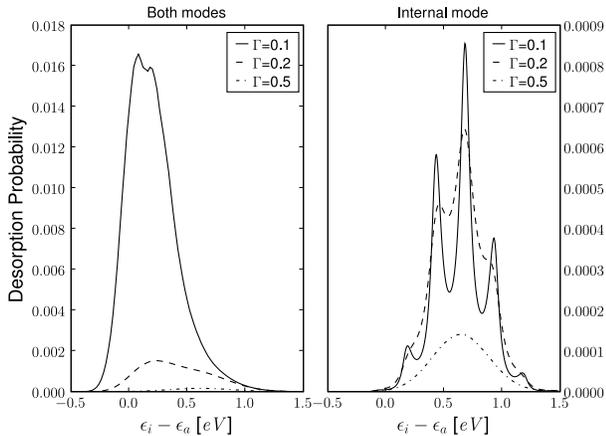} 
\caption{Desorption probability for CO adsorbed on Pt(111) for three different values of the resonance width. For $\Gamma>0.5$ the one and two-mode probability distributions becomes identical Lorentzians with an integrated probability that decays exponentially with resonance width (see Fig. \ref{fig:rate}).}
\label{fig:prob}
\end{figure}

In Fig. \ref{fig:prob} we display the probability that an incoming electron will scatter with an energy loss in excess of the desorption energy ($\Delta E>1.5 eV$) for three values of the resonance width. When only a single mode is considered we see the appearance of oscillator sidebands with an energy spacing of $\hbar\omega$. At larger resonance width the sidebands are washed out and the probability takes the form of a Lorentzian which is detuned by $\delta\epsilon_a\sim\Delta E/2$. A simple way to understand this detuning is as a compromise where both the incoming and outgoing electrons are closest to the resonance. Thus we see the emergence of an \textit{effective} inelastic resonance with a center that is detuned dependent on the desorption energy and a shape which is highly dependent on the lifetime. Such a probability distribution could not have been obtained in a model where the transfer of energy to the adsorbate is decoupled from the probability of capturing the electron, and the desorption probability would always be a Lorentzian (in the wide band limit) centered at $\epsilon_a$ and multiplied by a factor dependent on the details of the potential energy surfaces. For $\Gamma>0.5\;eV$ the COM degree of freedom becomes unimportant and the desorption probabilities obtained using both modes and only the internal degree of freedom become identical.

\begin{figure}[t]
  	\includegraphics[scale=0.4]{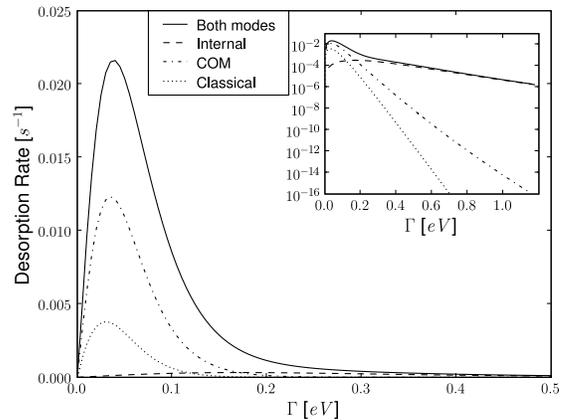} 
\caption{Desorption rate for CO adsorbed on Pt(111) as a function of resonance width $\Gamma$. In the wide resonance (short lifetime) regime the rate is seen to be completely governed by the internal stretch excitation whereas the COM excitation is governing the desorption rate in the narrow resonance (long lifetime) regime. The classical rate becomes several orders of magnitude smaller than the quantum rate at large resonance width. The inset shows the same data on a logarithmic scale.}
\label{fig:rate}
\end{figure}

Assuming an energy independent current of hot electrons we can integrate the desorption probabilities
in Fig. \ref{fig:prob} to obtain a desorption rate normalized to the incident flux of electrons. In Fig. \ref{fig:rate} we show how each of the two modes contribute to the desorption rate and compare with a calculation within the classical adiabatic model \eqref{clas}. 
The two single mode rates are obtained by setting $g_d$ and $g_z$ to zero in Eq. \eqref{P_mn}. It is seen that it is the internal stretch mode that governs the energy transfer completely in the large width regime and the COM mode governs the energy transfer at low width. 
The reason for this partitioning is the timescale associated with the two different modes. As seen from tables \ref{tab:par1} and \ref{tab:par2} the non-adiabatic coupling constants have approximately the same magnitude for the two modes. However, the period of oscillation is 5 times larger for the COM mode and for small lifetimes there is not enough time to transfer energy to the COM mode. From Fig. \ref{fig:rate} we see that the maximum rate of energy transfer in each mode occurs when $\Gamma\sim\hbar\omega_i$. The desorption rate decrease at small resonance width, since the hot electron then becomes weakly coupled to the resonant state

\begin{figure}[t]
  	\includegraphics[scale=0.4]{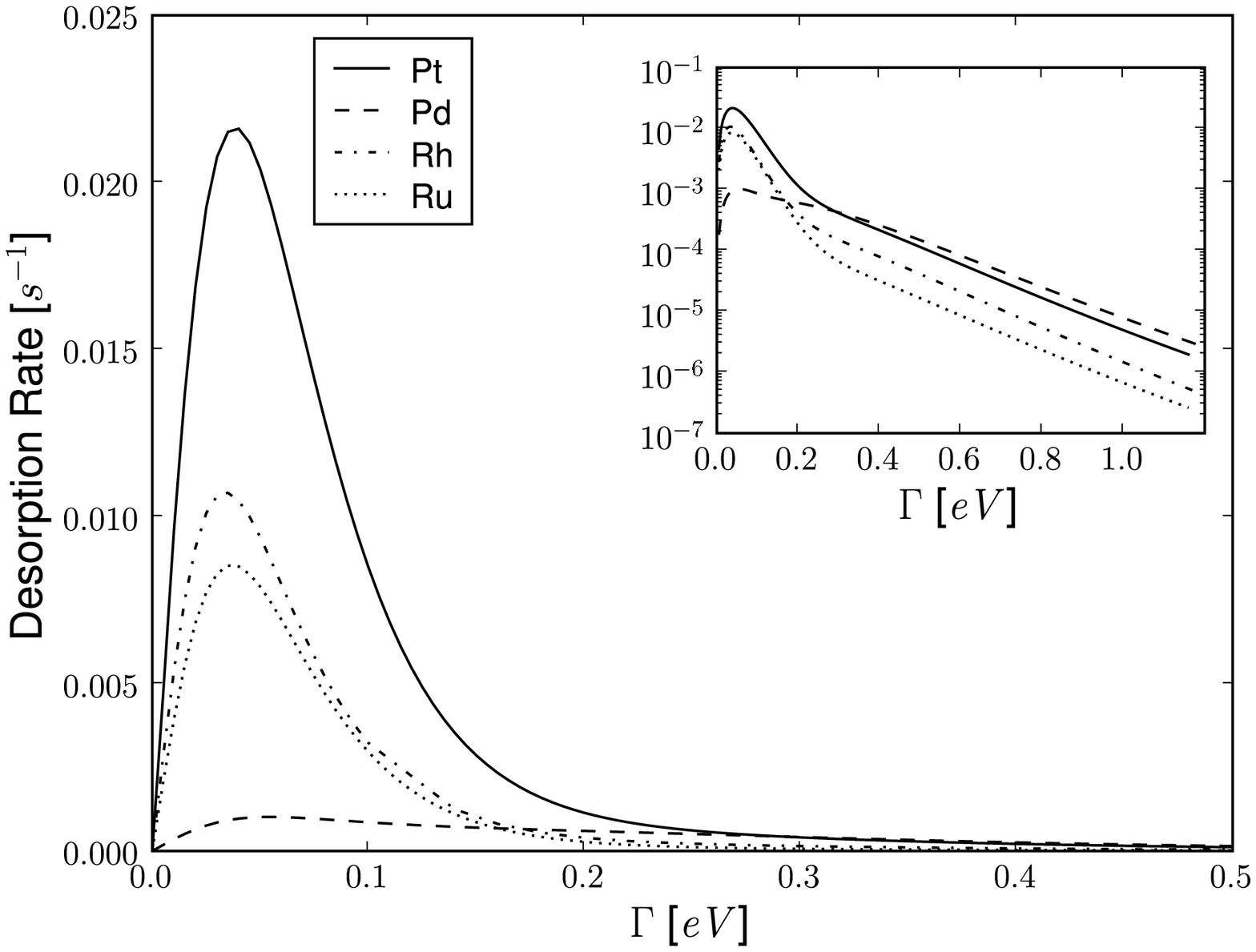} 
\caption{Rates of transferring $1.5\;eV$ to CO on 4 transition metals.}
\label{fig:COrates}
\end{figure}

\begin{figure}[t]
  	\includegraphics[scale=0.4]{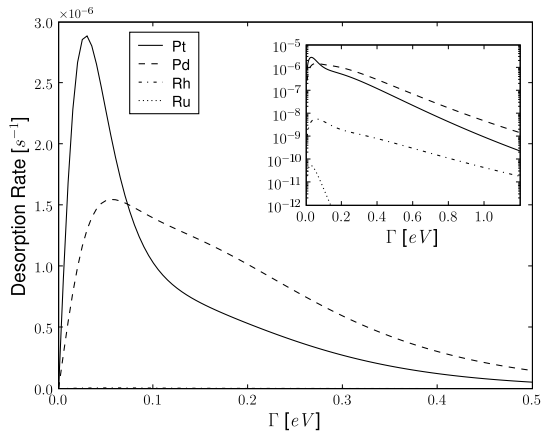} 
\caption{Rates of transferring $1.0\;eV$ to NO on 4 transition metals.}
\label{fig:NOrates}
\end{figure}

In Fig. \ref{fig:COrates} and \ref{fig:NOrates} we show a comparison of CO and NO adsorbed on the different transition metals. Again comparing with tables \ref{tab:par1} and \ref{tab:par2} it is seen that it is the coupling to the internal mode alone which controls the magnitude of the desorption rate at large resonance width. Since the internal degree of freedom seems to control the rate of energy transfer in the physical range of the resonance width (typically $0.5<\Gamma<1.5$) we will ignore the COM degree of freedom in the following. 

\subsubsection{Comparison of CO and NO}
So far we have analyzed some general features of desorption probabilities and their dependence on the non-adiabatic coupling parameters and the lifetime $\tau=\hbar/\Gamma$. Now we will compare the desorption probabilities of CO and NO on four transition metal surfaces using experimentally determined desorption energies. Although substantial experimental data exist for various systems including CO and NO, a direct comparison to experimental data is difficult since experimental desorption yields are highly dependent on the distribution of hot electrons in the substrate which depends on the detailed physical properties of the metal and the applied laser pulse. The distribution of hot electrons resulting from a given laser pulse could in principle be calculated from first principle, however, we will make no attempt of such a calculation here but simply compare desorption probabilities of single electron events as relevant for the MIM device\cite{gadzuk96, thomsen}. In tables \ref{tab:desorption1} and \ref{tab:desorption2} we summarize the desorption energy $E_d$, the estimated resonance width $\Gamma$, the detuning of the energy at which the incoming electron has the maximum probability of transferring the desorption energy $\delta\varepsilon = \varepsilon_i^{Max}-\epsilon_a$ and the maximum desorption probability $P_D^{Max} = P_D(\varepsilon_i^{Max})$ for the four transition metals (the maximum probability is detuned from $\epsilon_a$ as shown in Fig. \ref{fig:prob}).
\begin{table}[t]
\begin{center}
\begin{tabular}{c|c|c|c|c}
Metal & $E_D$ & $\Gamma$ & $\delta\varepsilon$ & $P_D^{Max}$ \\
	\hline
Pt(111)  & 1.37$^a$ & 1.0 & 0.6  & $2\cdot10^{-5}$ \\
Pd(111)  & 1.48$^a$ & 1.5 & 0.7  & $7\cdot10^{-7}$ \\
Rh(111)  & 1.45$^a$ & 1.2 & 0.7  & $1\cdot10^{-6}$ \\
Ru(0001) & 1.49$^a$ & 0.9 & 0.7  & $2\cdot10^{-6}$ \\
\end{tabular}
\caption{Desorption energies and calculated maximum desorption probability for CO adsorbed at top site on four transition metals. All numbers except $P_D^{Max}$ is $eV$. $^a$Experimental values taken from Abild-Pedersen and Andersson\cite{abild}.}
\label{tab:desorption1}
\end{center}
\end{table}
\begin{table}[t]
\begin{center}
\begin{tabular}{c|c|c|c|c|c}
Metal & $E_D$ & $\Gamma$ & $\delta\varepsilon$ & $P_D^{Max}$ \\
	\hline
Pt(111)  & 1.29$^b$ & 0.8 & 0.6 & $3\cdot10^{-11}$ \\
Pd(111)  & 1.17$^c$ & 0.6 & 0.6 & $5\cdot10^{-9}$  \\
Rh(111)  & 1.68$^c$ & 0.4 & 0.8 & $2\cdot10^{-15}$ \\
Ru(0001) & 1.49$^d$ & 0.3 & 0.7 & $3\cdot10^{-22}$ \\
\end{tabular}
\end{center}
\caption{Desorption energies and calculated maximum desorption probability for NO adsorbed at hcp hollow site on four transition metals. All numbers except $P_D^{Max}$ is $eV$. $^b$Croci et al\cite{croci}. $^c$Vang et al\cite{vang}, $^d$Butler et al\cite{butler}.}
\label{tab:desorption2}
\end{table}
The detuning very nicely follows the rule of thumb that $\delta\varepsilon\sim E_D/2$ in accordance with the picture of a compromise between the incoming and outgoing electron both being as close as possible to the center of the resonance $\epsilon_a$.

In general it is easier for a single electron at the right energy to mediate a desorption event involving CO than with NO from all the considered systems. However, in a femtosecond laser pulse experiment the resulting hot electron distribution would have much lower occupation numbers at the CO resonances than at a typical NO resonance. For example taking platinum as an example with a thermal electron distribution at $5000\;K$ and referring to tables \ref{tab:par1} and \ref{tab:par2} we see that the electronic occupation numbers at the resonance energy of CO and NO relates as $f(\epsilon_{NO})/f(\epsilon_{NO})\sim 150$. We should also note that the excited state potential energy surfaces for adsorbed NO is only quadratic in a small region near the minimum and the model Hamiltonian \eqref{h_quadratic} is thus not expected to describe NO as accurately as CO.

The desorption probabilities are highly dependent on the resonance width $\Gamma$ which we can only estimate roughly from the Kohn-Sham projected density of states. In addition, the electronic lifetime of CO on Pt(111) has been shown to be highly dependent on coverage\cite{fournier} since the $2\pi$ electrons becomes delocalized and quasistationary at certain coverages. Furthermore, both CO and NO are known to form adsorbate structures which is more involved\cite{vang, pedersen} than the simple periodic coverage of 0.25 monolayer considered here and the dependence of non-adiabatic coupling coefficients on coverage certainly deserves a study of its own. 

However, from Figs. \ref{fig:COrates} and \ref{fig:NOrates} we do observe the general trends that NO has a much weaker non-adiabatic coupling to the surfaces than CO and that for both CO and NO the coupling to Pt and Pd are similar whereas the coupling is weaker for Rh and very low for Ru. This decrease in non-adiabatic coupling could hint at a simple dependence on the number of d-band electrons. Investigating this will be the subject of future work.

\subsection{DIMET desorption rates}
To get an idea of desorption probabilities in the DIMET regime we will start by examining how an initial excitation influences the probability of transferring a given number of vibrational quanta. When the oscillator is in an excited vibrational state there is also the possibility of stimulated emission of vibrational quanta where the incoming hot electron gains energy by the scattering event.

\begin{figure}[t]
     \centering
%     \subfigure[$\Gamma=0.2$]{
%          \includegraphics[scale=0.12]{exc0.2.eps}
%     \subfigure[$\Gamma=0.5$]{
%          \includegraphics[scale=0.4]{exc0.5.eps}    
%     \subfigure[$\Gamma=1.0$]{
          \includegraphics[scale=0.4]{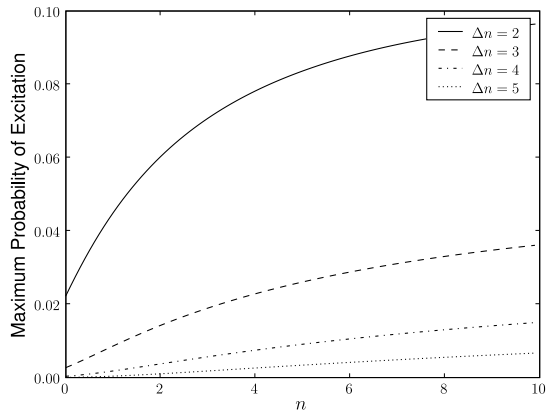}
\caption{Maximum probability of transferring $\Delta n$ vibrational quanta given that the initial state is $n$ with $\Gamma=1.0\;eV$.}
\label{dimet}
\end{figure}

In Fig. \ref{dimet} the maximum probability of transferring $\Delta n$ quanta is shown for a range of initial quantum numbers $n$. We treat $n$ as a continuous variable since in the case of a thermal ensemble of states the initial quantum number is simply replaced by a Bose distribution. There is a striking increase in the probabilities of transferring energy to the oscillator if the oscillator is already excited. For example, the probabilities of exciting $0\rightarrow3$ and $3\rightarrow6$ are $3\cdot10^{-3}$ and $2\cdot10^{-2}$ respectively although both transitions involve the same energy transfer. Thus if we compare the one-electron event $P_{0\rightarrow6}=6\cdot10^{-6}$ with the product of the two probabilities $P_{0\rightarrow3\rightarrow6}=6\cdot10^{-5}$ we get an order of magnitude difference and we still need to include the other channels for transferring 6 quanta in a two-electron event.

This also implies that the effect of a finite substrate temperature is two-fold: The occupation numbers of excited vibrational states will be non-zero meaning that less energy transfer is needed to desorb the molecule and the likelihood of a given energy transfer is increased if the molecule is thermally excited. However at room temperature the probability that the internal mode is in its first excited state is on the order of $~10^{-5}$ and we can safely neglect the effect of temperature.

A hallmark of the DIMET regime is the power law dependence of the desorption rate on the laser fluence $R\sim F^n$ where $n$ depends on the particular adsorbate/substrate system considered.\cite{prybyla1} It is by no means trivial that the desorption rate should follow a power law and calculating the exponent of a particular system is a major challenge of any DIMET model.

It is reasonable to that assume that the laser fluence is proportional to the flux of hot electrons hitting the molecule, since the desorption rate typically becomes linear\cite{ho} for small fluences corresponding to the DIET regime. As a simple model for the desorption rate we then consider a given flux $J$ of hot electrons at a fixed energy $\varepsilon_i$ hitting the resonance in equally spaced time intervals $\Delta t=1/J$. We assume that each vibrational quantum has a fixed lifetime $T_{vib}$ and that desorption occurs immediately if the vibrational energy reaches the desorption energy $E_D$. The probability that one vibrational quantum survives the time interval $\Delta t$ is $e^{-\Delta t/T_{vib}}$ and the probability of decay is $(1-e^{-\Delta t/T_{vib}})$. The probability that the first electron excites the $n$'th vibrational state is then simply the DIET probability
\begin{equation}
Q_1(n)=P_n(\varepsilon_i,0),
\end{equation}
where $P_n(\varepsilon_i,0)$ is given by Eq. \eqref{P_n}. The probability of the adsorbate being in the $n$'th vibrational state after the second electron has scattered is
\begin{equation}
Q_2(n)=\sum_{m=0}^{\infty}p(m)P_{n-m}(\varepsilon_i,m),
\end{equation}
where $P_{n-m}(\varepsilon_i,m)$ is the probability of the transition $m\rightarrow n$ (Eq. \eqref{P_n}) and $p(m)$ is the probability that the adsorbate was initially in the state $m$ given by
\begin{align}
p(m)=&\sum_{k=m}^{\infty}Q_1(k)\binom{k}{m}(e^{-\Delta t/T_{vib}})^m(1-e^{-\Delta t/T_{vib}})^{k-m}\notag\\
&\times\theta(E_d-\hbar\omega k).
\end{align}
Thus we only sum over values of $k$ below the desorption energy since states above $E_D$ would previously have been desorbed by assumption. Similarly the probability $Q_3(n)$ of being in the $n$'th excited state after the third scattering event can be expressed in terms of $Q_2(n)$ and so forth. The desorption probability of the $N$'th electron is then
\begin{equation}
P^{Des}_N=\sum_nQ_N(n)\theta(\hbar\omega n-E_D).
\end{equation}
When enough time intervals is included the probabilities converge such that $P^{Des}_N=P^{Des}_{N-1}$ and the desorption rate is $R(J)=J\cdot P^{Des}_N$ with $J=1/\Delta t$.

\begin{figure}[t]
     \centering
          \includegraphics[scale=0.41]{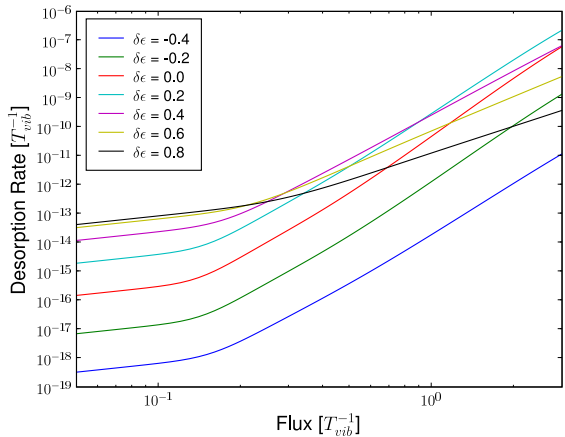}
\caption{Desorption rate as a function of electron flux per adsorption site. For small electron flux the rate is linear in the flux corresponding to the DIET regime whereas for larger electron flux the rate obeys a power law ($R\sim J^n$ with $n>1$) corresponding to the DIMET regime. In this figure we show the desorption rate of NO on Pt(111) using the parameters given in table \ref{tab:desorption2} and 7 different values of detuning.}
\label{fig:dimet_rate}
\end{figure}

In Fig. \ref{fig:dimet_rate} we show the rate for NO on Pt(111) with $\Gamma=0.8\;eV$. The desorption energy corresponds to 8 vibrational quanta. Note that changing the lifetime $T_{vib}$ in this model just corresponds to rescaling the flux. The similarity to similar experimental figures\cite{ho} is striking. At small flux the rate is linear whereas it obeys a power law ($R\sim J^n$ with $n>1$) at higher fluences. The fit to a power law is very good for fluxes above $0.2\;T_{vib}^{-1}$. For small values of the detuning ($-0.4<\delta\varepsilon<0.2\;eV$) we find that $5.5<n<6$  in good agreement with Ho\cite{ho}. For large positive values of the detuning the exponent decreases dramatically which is probably due the fact that fewer transitions dominate the dynamics in this region. This means that even though the results was obtained using the simple electron flux $J(\varepsilon_i)=J_0\delta(\varepsilon_i-\epsilon_a-\delta\varepsilon)$ we would most likely obtain the same exponent if we generalized the model to any flux localized within $\pm0.2\;eV$ of the resonance. 

Although the correspondence with the experimentally found exponent may be fortuitous in such a simple model the power law itself is very robust to changes in the parameters and we obtain similar power laws for CO on Pt(111). For example, changing the value of $\Gamma$ results in an overall shift of the rates but the exponents are essentially unchanged. Indeed the exponents appear to be determined mainly by the number of vibrational quanta needed for desorption.

\section{Summary and discussion}
We have previously presented a method to obtain excited state potential energy surfaces for molecules chemisorbed at metal surfaces.\cite{gavnholt} In this paper the method have been applied and combined with a non-adiabatic quantum model to obtain desorption probabilities for CO and NO on four transition metal surfaces.

The model we have applied allow us to predict the probability that a hot electron will transfer a given amount of energy to the different vibrational modes of an adsorbate. Our main conclusion is the significant role of the internal degree of freedom and the failure of classical mechanics to describe the excited state adsorbate propagation. 

Combining the model with a simple picture of the decay and re-excitation of vibrational states reproduces the characteristic power laws of DIMET experiments and yields the exponent associated with a given adsorbate/substrate system. 

The model we have used for calculating the energy transfer rates obviously represents a very simplified view of the dynamics. First of all it is a model of non-interacting electrons. We assume that we can include the important part of the electron-electron interactions by using non-adiabatic coupling coefficients $\lambda_i$ obtained from the interacting density with linear expansion $\Delta$SCF-DFT. The approximation amounts to assuming ballistic hot electrons and instantaneous restructuring of the electronic environment when occupying the resonance. Although this may be the case in some metallic systems electron-electron interactions could have effects which go beyond a simple renormalization of the non-adiabatic coupling. The linear non-adiabatic coupling regime leading to Eq. \eqref{h_quadratic} corresponds to an assumption of equal curvature on the ground and excited state PES. This is a good approximation for CO but NO has a very shallow excited state PES on some of the transition metals and there the approximation may not be as good.

Furthermore the model assumes that the ground state potential is quadratic and that the excited state potential is simply a shifted ground state potential. At least in the COM direction it is clear From Fig. \ref{fig:pes} that the ground state potential deviates significantly from a quadratic potential and since we are concerned with high lying vibrational excitations this deviation could perhaps have an important effect. It may be possible to include anharmonic terms in the Hamiltonian and calculate new scattering amplitudes perturbatively but the this will be left for future work.

We have focused on the molecules CO and NO, since they have a
conceptually simple structure and a vast amount of experiments have
been performed on these systems. However, it is well known that
GGA-DFT calculations of CO adsorbed an Pt(111) predicts CO to bind at
a hollow site in contradiction to the experimentally observed top
site\cite{feibelman}.  While the difference in adsorbtion energy
appears to be less with \texttt{gpaw} than in the calculations
presented in Ref.~\onlinecite{feibelman}, possibly due to the use of
the PAW method instead of ultrasoft pseudopotentials, the difference
is still 80 meV and the inability to predict the correct binding site
is worrying.  On the other hand, the existence in the calculation of
another adsorption site with a slightly lower energy is unlikely to
change the local shape of the potential energy surface enough to
\emph{qualitatively} change the results obtained here.  In addition,
we see a very similar behavior for CO on Ru(0001), where DFT does
predict the right adsorption site (the top site).  We have thus chosen
to put CO at the experimentally observed top site as the hollow site
would lead to a smaller surface molecule distance and thus very
different screening and desorption rate.

As previously mentioned the value of $\Gamma$ is estimated from the Kohn-Sham projected density of states, but we do not know how well this estimate matches the true value and as such we have mostly treated $\Gamma$ as a free parameter. In fact the object of interest in the problem is the spectral function of the resonant state, but even if we had a reliable way of determining this function we would have to make the wide band approximation (where the spectral function is a Lorentzian of width $\Gamma$) in order to calculate scattering rates. Nevertheless it would be very interesting to calculate this function to get an idea of the validity of the wide band approximation and to obtain a trustworthy value of $\Gamma$.

We have not made any attempt to predict how the distribution of energy evolves after a molecule returns to its electronic ground state, but assume that the dissipation of energy is slow enough that the adsorbate will desorb if the desorption energy has been transferred. This is of course a rather crude assumption and the rate of energy transfer should be accompanied by a detailed molecular propagation on the full dimensional ground state PES to improve the results. Ground state molecular dynamics would also be necessary to obtain branching ratios when there is a possibility of different chemical reactions induced by hot electrons. 

However the model we have presented captures some of the essential features of non-adiabatic dynamics. For example the appearance of an effective inelastic resonance which is detuned from the electronic resonance by an amount depending on the energy transfer is a pure non-adiabatic result and would never have emerged from an adiabatic model. Furthermore the exponents in the DIMET power laws appear to be determined by the number of vibrational quanta needed for desorption and thus communicates the quantum nature of the dynamics.

\appendix
\section{Projecting KS states on a molecular orbital in PAW}\label{paw}
The Projector Augmented Wave (PAW) method\cite{blochl1} utilizes that one can transform single-particle wavefunctions $|\psi_n\rangle$ oscillating wildly near the atom core (all-electron wavefunctions), into smooth well-behaved wavefunctions $|\tilde\psi_n\rangle$ (pseudo wavefunctions) which are identical to the all-electron wavefunctions outside some augmentation sphere. The idea is to expand the pseudo wavefunction inside the augmentation sphere on a basis of smooth continuations $|\tilde\phi_i^a\rangle$ of partial waves $|\phi_i^a\rangle$ centered on atom $a$. The transformation is
\begin{align}\label{PAW}
|\psi_n\rangle=|\tilde\psi_n\rangle+\sum_{i,a}\Big(|\phi_i^a\rangle-|\tilde\phi_i^a\rangle\Big)\langle\tilde p_i^a|\tilde\psi_n\rangle,
\end{align}
where the projector functions $|\tilde p_i^a\rangle$ inside the augmentation sphere $a$ fulfills
\begin{align}\label{dual}
\sum_i\langle\tilde p_i^a|\tilde\phi_i^a\rangle=1,\qquad\langle\tilde p_i^a|\tilde\phi_j^a\rangle=\delta_{ij},\qquad|\mathbf{r}-\mathbf{R}^a|<r^a_c.\notag
\end{align}

\begin{figure}[t]
	\includegraphics[scale=0.3]{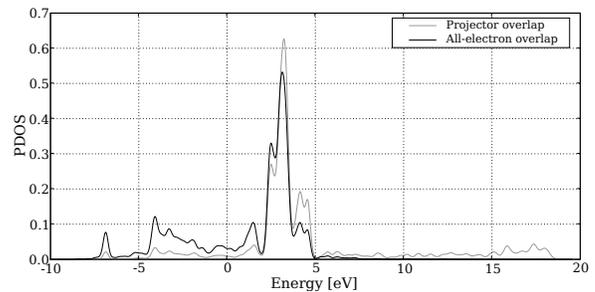}  
\caption{Ground state calculation of CO adsorbed on Pt(111) top site. The projected density of states of the $2\pi$ orbitals using the methods of projector/pseudo wavefunction overlap and all-electron wavefunction overlap are compared. In the projector overlap method the orbital is defined by $|\tilde{p}_{2\pi}\rangle=\frac{1}{\sqrt{13}}\big(3|\tilde{p}_x\rangle_C-2|\tilde{p}_x\rangle_O\big)$ which is the orbital most similar to the gas-phase calculation. The long high energy tail of the projector overlap signals an inaccuracy of the method and makes excited state calculations dependent on the number of unoccupied bands. Thus we use the all-electron overlaps to determine expansion coefficients in this work.}
	\label{fig:pdos}
\end{figure}

The method of linear expansion $\Delta$SCF involves expanding a molecular orbital $|\varphi_i\rangle$ in Kohn-Sham states $|\psi_n\rangle$ and do a self consistent calculation with an additional density corresponding to the orbital.\cite{gavnholt}. The simplest way of getting the expansion coefficients is using the projector overlaps $\langle\psi_n|\varphi_i\rangle\sim\langle\tilde\psi_n|\tilde p_i^a\rangle$ which is calculated in each iteration anyway. However, this method turns out to be too inaccurate in the case of CO on Pt(111) due to non-vanishing projector overlaps for highly energetic Kohn-Sham states as shown in Fig. \ref{fig:pdos}. This implies that the expansion coefficients depend on the number of unoccupied bands included in the calculation

To calculate the overlaps $\langle\psi_n|\varphi_i\rangle$ exactly, one should start by performing a gas-phase calculation of the molecule or atom which is to be used
in the $\Delta$SCF calculation. The pseudo wavefunction $\tilde\psi_i(x)$ corresponding to the orbital to be occupied is then saved along with the projector overlaps $\langle\tilde p_k^a|\tilde\psi_i\rangle$ and the $\Delta$SCF calculation is initialized. In each step of the calculation we can then do a numerical integration to obtain the expansion coefficients by
\begin{align}
c_{ni}&=\langle\psi_n|\psi_i\rangle \\
&=\langle\tilde\psi_n|\tilde\psi_i\rangle+\sum_{a,j,k}\langle\tilde\psi_n|\tilde p_j^a\rangle\Big(\langle\phi^a_j|\phi_k^a\rangle-\langle\tilde\phi^a_j|\tilde\phi_k^a\rangle\Big)\langle\tilde p_k^a|\tilde\psi_i\rangle,\notag
\end{align}
where \eqref{PAW} was used. Note that there is only a single sum over atoms (and only the ones in the molecule) and that the cross terms of pseudo/all-electron wavefunction does not contribute. This can be seen using the arguments following Eq. 20 in Ref. \onlinecite{blochl2}. 

\section{Calculating the inelastic scattering probability}\label{prob}
Here we briefly summarize the calculation leading to the inelastic scattering probabilities in the model \eqref{h_quadratic}.\cite{wingreen}
An explicit expression for the probability has previously been obtained\cite{wingreen} for a single mode at initially in the ground state. Here we will extend the result to an explicit expression for any number of modes initially in a thermal ensemble of vibrationally excited states.

From the Hamiltonian \eqref{h_non-ad} the differential reflection matrix $R(\varepsilon_i,\varepsilon_f)$ which is defined as the probability per unit final state energy that an incoming hot electron with energy $\varepsilon_i$ scatters on the resonance into a final state of $\varepsilon_f$, can be expressed in terms of a four point Green function. The inelastic part is contained in the expression:
\begin{widetext}
\begin{align}\label{r_matrix}
R_{in}(\varepsilon_i,\varepsilon_f)=\Gamma(\varepsilon_f)\Gamma(\varepsilon_i)\int\int\int\frac{d\tau ds dt}{2\pi\hbar^3}e^{i[(\varepsilon_i-\varepsilon_f)\tau+\varepsilon_ft-\varepsilon_is]/\hbar}G(\tau,s,t),
\end{align}
where the Green functions is
\begin{equation}
 G(\tau,s,t)=\theta(s)\theta(t)\langle c_a(\tau-s)c_a^\dag(\tau)c_a(t)c_a^\dag(0)\rangle,\qquad c(t)=e^{iHt/\hbar}c(0)e^{-iHt/\hbar},
\end{equation}
and $\langle\rangle$ denotes a thermal ensemble of oscillator states. The expression is valid for any non-adiabatic coupling function $\varepsilon_a(x)$, but in general it can be very hard to obtain an expression for the Green function. An exception is the wide band limit with linear coupling corresponding to the Hamiltonian \eqref{h_quadratic}.\cite{wingreen} The Green function then becomes
\begin{equation}\label{green}
G(\tau,s,t)=\theta(t)\theta(s)e^{-i\epsilon_a(t-s)/\hbar-\Gamma(t+s)/2\hbar}\exp\bigg(\sum_ig_i\Big[i(t-s)\omega_i-(1+n_i)f_i-n_if_i^*\Big]\bigg),
\end{equation}
where $\epsilon_a$ is center of the resonance, $n_i$ is the Bose distribution, $g_i$ = $(\lambda_i/\hbar\omega_i)^2$ is the effective coupling constant of the mode $i$, and
\begin{equation}
f_i(\tau,s,t)=2-e^{-i\omega_it}-e^{i\omega_is}+e^{-i\omega_i\tau}(1-e^{i\omega_it})(1-e^{i\omega_is}).
\end{equation}
The integrals in the scattering matrix \eqref{r_matrix} can be evaluated by writing the exponentials in \eqref{green} as Taylor expansions and performing the $\tau$ integral. This leaves the remaining two integrals as complex conjugates which are evaluated by writing factors such as $(1-e^{i\omega_it})^m$ by their binomial expansions. For a single oscillator with thermal occupation $n$ we obtain the inelastic reflection matrix:
\begin{align}\label{P_T}
R_{in}(\varepsilon_i, \varepsilon_f, n)&=\Gamma^2e^{-2g(1+2n)}\sum_{m_1=0}^{\infty}\sum_{m_2=0}^{\infty}\frac{g^{m_1+m_2}(1+n)^{m_1}n^{m_2}}{m_1!m_2!}\delta\big(\varepsilon_i-\varepsilon_f-(m_1-m_2)\hbar\omega\big)\times F(m_1,m_2),
\end{align}
with
\begin{align}
F(m_1,m_2)=\Bigg|&\sum_{i=0}^{m_1}\sum_{j=0}^{m_2}(-1)^{i+j}\binom{m_1}{i}\binom{m_2}{j}\sum_{k=0}^\infty\sum_{l=0}^\infty\frac{g^{k+l}(1+n)^kn^l}{k!l!}
\frac{1}{\varepsilon_i-\epsilon_a-(i-j+k-l-g)\hbar\omega+i\Gamma/2}\Bigg|^2.\notag
\end{align}
Although the expression looks rather complicated it has a simple interpretation. Integrating over final state energies in the vicinity of $\Delta n=m_1-m_2$ gives the probability of transferring $\Delta E=\Delta n\hbar\omega$ to the oscillator if the energy of the incoming electron is $\varepsilon_i$:
\begin{align}\label{P_n}
P_{\Delta n}(\varepsilon_i, n)=\Gamma^2e^{-2g(1+2n)}\Big(&\frac{g^{\Delta n}(1+n)^{\Delta n}}{\Delta n!}F(\Delta n,0)+\frac{g^{\Delta n+1}(1+n)^{\Delta n+1}gn}{(\Delta n+1)!}F(\Delta n+1,1)\\
&+\frac{g^{\Delta n+2}(1+n)^{\Delta n+2}(gn)^2}{(\Delta n+2)!2!}F(\Delta n+2,2)+\ldots\Big)\notag,
\end{align}
where the first term is the probability of adding $\Delta n$ bosons, the second term is the probability for removing (coupling $ng$) one and adding (coupling $(n+1)g$) $\Delta n+1$ bosons and so forth. 

We can also evaluate the differential reflection matrix for $N$ oscillators initially in the ground state with frequencies and coupling constants $\omega_i$ and $g_i$ respectively. The result is
\begin{align}\label{P_i}
R_{in}(\varepsilon_i, \varepsilon_f)&=\Gamma^2e^{-2\sum_{i=1}^Ng_i}\sum_{m_1=0}^{\infty}\ldots\sum_{m_N=0}^{\infty}\frac{g_1^{m_1}\ldots g_N^{m_N}}{m_1!\ldots m_N!}\delta\big(\varepsilon_i-\varepsilon_f-\sum_{i=1}^Nm_i\hbar\omega_i\big)\\
&\times\Bigg|\sum_{j_1=1}^{m_1}\ldots\sum_{j_N=1}^{m_N}(-1)^{\sum_{i=1}^Nj_i}\binom{m_1}{j_1}\ldots\binom{m_N}{j_N}\sum_{l_1=0}^\infty\ldots\sum_{l_N=0}^\infty\frac{g_1^{l_1}\ldots g_N^{l_N}}{l_1!\ldots l_N!}\cdot
\frac{1}{\varepsilon_i-\epsilon_a+i\Gamma/2-\sum_{i=1}^N(j_i+l_i-g_i)\hbar\omega_i}\Bigg|^2.\notag
\end{align}
It is amusing that the result \eqref{P_T} for a one mode system with initial excitation number $n$ follows from the result \eqref{P_i} if we regard \eqref{P_T} as a two-mode system at $T=0$ with energies $\hbar\omega$ and $-\hbar\omega$ and coupling constants $g(n+1)$ and $gn$ respectively. For convenience we state the probability of exciting the $(m_d,m_z)$ state from the ground state in the two-dimensional model with modes $d$ and $z$
\begin{align}\label{P_mn}
P_{m_dm_z}(\varepsilon_i)&=\Gamma^2e^{-2(g_d+g_z)}\frac{g_d^{m_d}g_z^{m_z}}{m_d!n_z!}\\
&\times\Bigg|\sum_{j_d=1}^{m_d}\sum_{j_z=1}^{m_z}(-1)^{j_d+j_z}\binom{m_d}{j_d}\binom{m_z}{j_z}\sum_{k=0}^\infty\sum_{l=0}^\infty\frac{g_d^kg_z^l}{k!l!}\cdot
\frac{1}{\varepsilon_i-\epsilon_a-(j_d+k-g_d)\hbar\omega_d-(j_z+l-g_z)\hbar\omega_z+i\Gamma/2}\Bigg|^2\notag
\end{align}

\subsection{Elastic scattering}
The elastic part of the scattering matrix for a single oscillator with thermal occupation number $n$ is
\begin{align}\label{P_el}
R_{el}(\varepsilon_i, \varepsilon_f, n)=&\delta(\varepsilon_i-\varepsilon_f)(1+2\text{Im}G_R(\varepsilon_i)),\qquad G_R(\varepsilon)=\int\frac{dt}{\hbar}e^{i\varepsilon t/\hbar}G_R(t),\\
&\qquad G_R(t)=-i\theta(t)\langle n|c_a(t)c_a^\dag(0)|n\rangle.\notag
\end{align}
We can use the linked cluster theorem to derive the retarded Green function and get the result
\begin{align}
G_R(t)=-i\theta(t)e^{-g(1+2n)}e^{(-i\epsilon_a-ig\hbar\omega -\Gamma/2)t/\hbar}\sum_{m_1=0}^\infty\sum_{m_2=0}^\infty\frac{g^{m_1}n^{m_1}g^{m_2}(1+n)^{m_2}}{m_1!m_2!}e^{-i(m_2-m_1)\omega t}
\end{align}
We can then calculate the elastic part of the scattering probability and get
\begin{equation}
P_{el}(\varepsilon_i,n)=1-\Gamma^2e^{-g(1+2n)}\sum_{m_1=0}^\infty\sum_{m_2=0}^\infty\frac{g^{m_1}n^{m_1}g^{m_2}(1+n)^{m_2}}{m_1!m_2!}\cdot\frac{1}{(\varepsilon_i-\epsilon_a-[m_2-m_1-g]\hbar\omega)^2+(\Gamma/2)^2}.
\end{equation}
\end{widetext}
When calculating the elastic scattering probability one should also remember to include the $m_1=m_2$ terms in \ref{P_T}.

The $n$ in the expressions above denote the Bose distribution and not a specific state $|n\rangle$, but in the context of DIMET our main point of interest is the probability that a oscillator initially in the state $|n_i\rangle$ scatters inelastically to the state $|n_f\rangle$. However, the expression in the case of a pure state is very similar to the thermal ensemble, the only difference being that we should make the substitution
\begin{equation}
e^{-g_in_i(f_i+f_i^*)}\rightarrow L_{n_i}(g(f_i+f_i^*))
\end{equation}
in \eqref{green}, where $L_n(x)$ is the $n$'th Laguerre polynomial. The expression involving Laguerre polynomials is somewhat more complicated to handle numerically and therefore we have chosen to work with the thermal ensemble expressions instead. In the range of parameters in the present work the thermal ensemble expressions are very good approximations since $L_n(x)$ have the same first order Taylor expansion as $e^{-nx}$ and for $t<\hbar/\Gamma$ we get $g_if_i< 0.001$.

\begin{acknowledgments}
We would like to acknowledge the help and advise of Jens J\o rgen Mortensen on implementing and using linear expansion $\Delta$SCF in gpaw. The Center for Individual Nanoparticle Functionality (CINF) is sponsored by the Danish National Research Foundation. This work was supported by the Danish Center for Scientific Computing.
\end{acknowledgments}

\newpage

%\bibliography{bibfile}{}

\end{document}